\documentstyle[epsf]{article}
\begin{document}
\renewcommand{\thefootnote}{\fnsymbol{footnote}}
\begin{flushright}
KEK-preprint-94-71\\
\end{flushright}
\vskip -1cm
%\special{psfile=kekm.epsf hscale=0.7 vscale=0.7 }
\epsfysize3cm
\epsfbox{kekm.epsf}
\begin{center}
{\Large \bf
New Tagging Method of B Flavor of Neutral B Meson \\
in CP Violation Measurement \\
in Asymmetric B-Factory Experiment
\footnote{
published in Journal of the Physical Society of Japan
{\bf Vol. 63}, No. 10, Oct., 1994, pp. 3542-3545.
}
}\\
\vskip 0.5cm
Ryoji Enomoto
\footnote{Internet address: enomoto@kekvax.kek.jp}\\
\vskip 0.5cm
{\it
National Laboratory for High Energy Physics, KEK,\\
1-1 Oho, Tsukuba, Ibaraki-ken 305, Japan
}\\
\end{center}
\begin{abstract}
In CP violation measurements in asymmetric B-factory experiments,
a determination of the B flavor of the neutral B mesons is necessary.
A new method
to this purpose
using only three vectors of charged particles has been developed.
This method (weighted charge
method) does not require either lepton identification or
charged-kaon
identification. The tagging efficiency, probability for incorrect
tagging, and effective
tagging efficiency of this method
are 43.1, 18.3, and 17.3\%, respectively.
\end{abstract}
\vskip 0.5cm
%introduction
%CP violation measurements represent a major topic concerning
%asymmetric
%B-Factory experiments\cite{belle,babar}.
Asymmetric B-factories are being constructed to measure CP
violation parameters
in the B system\cite{belle,babar}.
In such measurements, for example,
$B^0(\overline{B^0})\rightarrow \psi K_s$ decay must
be identified by a full reconstruction. One should thus identify the
B flavor of the other $B^0(\overline{B^0})$ meson
[$B^0(\overline{B^0})$-tagging]
in order to determine the CP eigenstate
at its decay timing.
Hereafter, the description of $B^0$ includes its charge conjugation,
i.e., $\overline{B^0}$.
The typical methods of $B^0$-tagging are
lepton and charged-kaon tagging\cite{aleksan}.
The effective tagging efficiency is written as
$$\epsilon_{eff}=\epsilon (1-2w)^2,$$
where $\epsilon$ and $w$ are tagging efficiency and the probability
for incorrect tagging, respectively.
A typical value is $\epsilon_{eff}$=0.28\cite{belle}.
However, these methods require expensive devices, such as a CsI calorimeter
and a ring imaging \v Cerenkov detector (RICH) \cite{rich}.
Moreover, the performance of the RICH is still unknown,
for example regarding the quantum efficiencies of the CsI
photocathodes\cite{enomoto}.
In this report, a new method (weighted charge method)
which does not require either lepton
or charged-kaon identification is introduced.

%two body decay
The dominant decay topology of the $B^0$ meson is
back-to-back $(W)\rightarrow q\bar{q},l\bar{l}$ and charm jets,
where $(W)$ is an intermediate $W$ boson.
Charm jets are also back-to-back $(W)$ and strange jets in the center
of the mass frame of the charmed particle.
There is therefore a high probability that the charge state of the
highest-momentum particle represents the B flavor of the parent $B$ meson.
Also, the kaon from the $D$ meson may have a higher momentum because
the $D$ meson is boosted.

%simple test
The present work began from highest-momentum particle tagging.
The detector geometry assumed in this study is similar to
that of the KEK
B-factory experiment\cite{belle}.
The beam energies were assumed to be 3.5 and 8 GeV. The detector
covers polar-angle ($\theta$) regions between 17 and 150 degrees.
For a $B$-decay generator, JETSET6.3 was used\cite{lund63}.
A $B^0$ meson was generated along the beam-axis (zero polar angle)
with an energy of 5.75 GeV. This is because
in the case of $B^0\rightarrow \psi
K_s$ decays,
particles from this $B^0$ meson can be totally removed from the track
sample. Also, the four-vector of the $B^0$ meson is exactly known at this
point.
Then, cuts at the polar-angle coverage and for $P_T>$ 0.15 GeV
were subjected to the decay products of the $B^0$.
The energies of all the charged particles were calculated
assuming that they were charged pions.
Finally, they were boosted to the CMS frame of this $B^0$ meson.
In this method, $\epsilon$ is 100\% by definition,
because there is always a highest-momentum particle.
A $w$ was obtained to be 33\%.
Thus, $\epsilon_{eff}$ of 12\% was obtained. This value is
comparable to that of lepton-tagging.

\begin{figure}
%\vskip 7cm
\vskip -2cm
\epsfysize10cm
%\special{psfile=fig1.eps hscale=0.8 vscale=0.8 voffset=-216}
\hskip1in\epsfbox{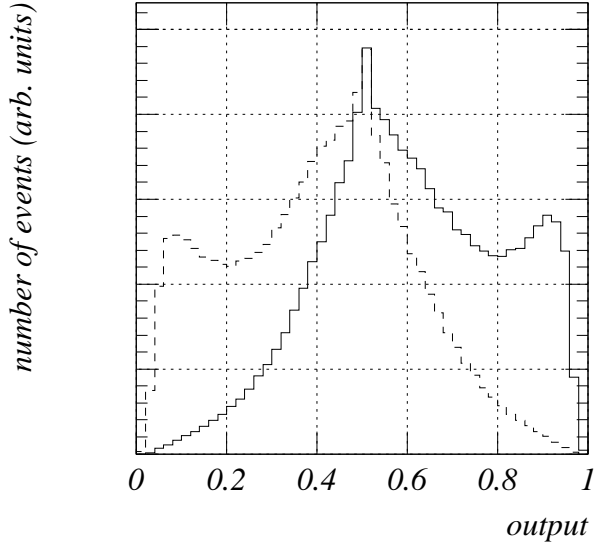}
\vskip -1cm
\caption{Output distributions of the neural-network.
The horizontal axis is the output of the neural network and the
vertical is the number of events.
The solid histogram is that for
$B^0$ and the dashed one is that for $\overline{B^0}$.}
\label{jn}
\end{figure}

\begin{figure}
%\vskip 7cm
%\special{psfile=fig2.eps hscale=0.8 vscale=0.8 voffset=-216}
\vskip -2cm
\epsfysize10cm
\hskip1in\epsfbox{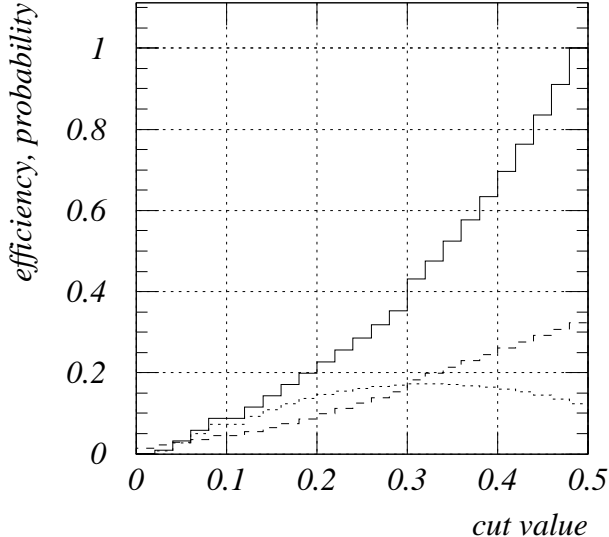}
\vskip -1cm
\caption{Cut dependences of $\epsilon$, $w$, and $\epsilon_{eff}$.
The description of ``cut" is described in the text.
The solid histogram is for $\epsilon$, the dashed for $w$, and the dotted
for $\epsilon_{eff}$.}
\label{cut}
\end{figure}

%neural-network
In order to determine the other signatures, such as the $D$-decay topologies,
a neural-network program (JETNET1.1) was used\cite{jetnet}.
The input parameters were the four highest three-vectors and their charges.
One middle layer with ten parameters was assumed.
This program (neural network) calculates a quantity (we simply call it
``output") which is related to the B flavor, i.e.,
output=0 means B=-1 and output=1 means B=+1. A cut was
applied on this quantity in order to determine the B flavor.
The output of the neural-network is shown in Figure \ref{jn}.
The solid histogram is that for
$B^0$ and the dashed is that for $\overline{B^0}$.
The cut dependences on $\epsilon$, $w$, and $\epsilon_{eff}$
are shown in Figure \ref{cut} by the solid, dashed, and dotted histograms,
respectively.
If cuts were made
at 0.69 and 0.31 for $B^0$ and $\overline{B^0}$, respectively,
$\epsilon$ and $w$ of 43.1 and 18.3\% were obtained,
respectively.
Thus, $\epsilon_{eff}$ was obtained to be 17.3\%.
This value is half that in the case of lepton and charged-kaon tagging.
Hereafter, $\overline{cut}$ is used for the cut value of the
neural-network output for $\overline{B^0}$.

So far we assumed perfect momentum resolution.
The momentum resolution dependence was studied. Here, a resolution of
$$\delta P_T/P_T = \sqrt{0.005^2+(0.005P_T)^2}$$
was assumed. $\epsilon$, $w$, and
$\epsilon_{eff}$ of 43.2, 18.2, and 17.5\% at $\overline{cut}$=0.31
were obtained. Therefore, no deterioration due to the momentum resolution
was observed.

%lund7.3
Since the decay-mode table
of charmed particles in JETSET6.3 is not perfect, the decay-mode
dependence should be verified. For this, JETSET7.3
was used\cite{lund73}.
An $\epsilon_{eff}$ of 17.4\% with $\epsilon$=47.8\% and $w$=19.9\%
was obtained at a similar $\overline{cut}$
level of 0.35.
The ambiguity due to the decay modes of
the charmed particles was ensured to be small.
In the following studies,
JETSET6.3 was used for this reason.

%overlap with lepton- and kaon-tagging
Considering realistic conditions of B-factory experiments,
lepton identification at $P^*>$1.4 GeV
was assumed to be perfect,
where $P^*$ is the lepton momentum in the $B^0$ rest frame.
In Reference \cite{aleksan}, the lepton tagging efficiency is
14\% with 6\% for incorrect tagging. In the KEK B-factory letter of intent
\cite{belle}, the lepton tagging efficiency is 10.6\% with 4\%
for incorrect
tagging. In our simulation, the lepton identification probability was
assumed to be 100\%. Then, an $\epsilon$ of 12.1\% with $w$=3.3\%
was obtained, assuming the above $P^*$ cut.

In addition, charged-kaon tagging was considered. There are two options
for an asymmetric B-factory,
i.e. (a) with and (b) without the RICH.
For case (a), the identification of the kaon was assumed to be perfect at
$P<$3.5 GeV; for case (b), $P<$1.0 GeV.
Also, a special case (c) was added, i.e., a no-kaon identification option.
This option is for a poor-man's hadron B-factory\cite{bcd}.
For case (a), $\epsilon$ and $w$ were obtained to be 39.2 and 8.5\%
and for case (b), they were 30.4 and 10.6\%, respectively.
Using both lepton and charged-kaon tagging
and considering their overlap,
$\epsilon_{eff}$'s of 34.0 and 26.6\% were obtained for cases (a) and
(b), respectively.

\begin{figure}
%\vskip 10cm
%\special{psfile=fig3.eps hscale=0.8 vscale=0.6 voffset=-18}
\epsfysize20cm
\hskip1in\epsfbox{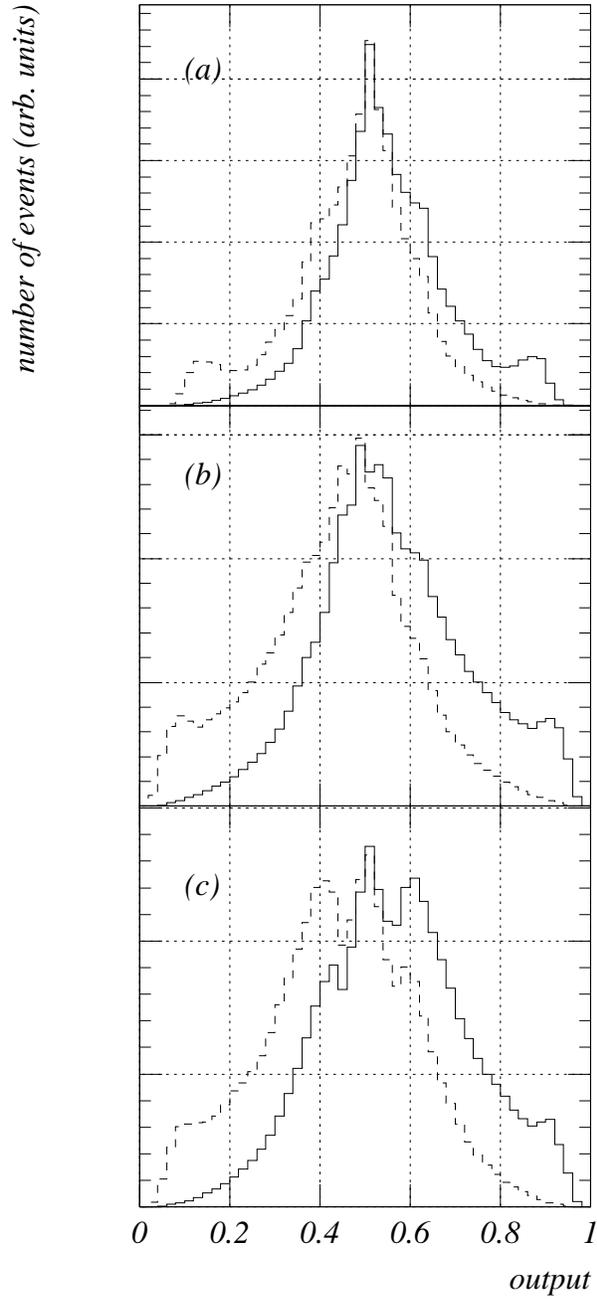}
\caption{Output distributions of the neural-network studies: (a), (b), and (c)
correspond to the cases described in the text.
The horizontal axis is the output of the neural network and the
vertical is the number of events.
The solid histograms are
for $B^0$ and the dashed $\overline{B^0}$.}
\label{feff}
\end{figure}

The results of the weighted charge method were as follows.
The events tagged by either the lepton or the charged-kaon were
removed from the samples.
Thus, the efficiencies mentioned hereafter were
re-defined considering a reduction of an event portion by the
lepton and/or charged kaon taggings.
The results are summarized in Table \ref{teff}.
%\onecolumn
\begin{table}
\begin{center}
\begin{tabular}{ccccc}
\hline
\hline
 &\multicolumn{4}{c}{lepton and kaon tags}\\
Case & lepton-ID & kaon-ID & $\epsilon^0_{eff}$ & \\
 & $P^*$ cut & $P$ cut &&\\
\hline
(a) & 1.4 GeV  & 3.5 GeV & 33.7& \\
(b) & 1.4 GeV & 1.0 GeV & 26.5 &\\
(c) & 1.4 GeV & none& 10.4 &\\
(d) & none& none& 0    &\\
\hline
 & \multicolumn{4}{c}{this method} \\
Case & $\epsilon$
&$w$ & $\epsilon_{eff}$ & $\epsilon_{eff}^{tot}$ \\
\hline
(a) & 18.1 & 29.2 & 3.12 & 36.8 \\
(b) & 24.4 & 24.3 & 6.42 & 32.9 \\
(c) & 43.2 & 27.1 & 9.03 & 19.4 \\
(d) & 43.1 & 18.3 & 17.3 & 17.3 \\
\hline
\hline
\end{tabular}
\end{center}
\caption{Percentages of $\epsilon$, $w$, and $\epsilon_{eff}$
for four cases described in the text.
Here, the efficiencies were renormalized by excluding the lepton-kaon
tagging events as described in the text.
Also shown are the effective tagging efficiencies for the
lepton-kaon tagging ($\epsilon_{eff}^0$)
and total effective tagging efficiencies, i.e.,
$\epsilon_{eff}^{tot}=\epsilon_{eff}^0+\epsilon_{eff}$,
respectively.
}
\label{teff}
\end{table}
%\twocolumn
Case (d) in Table \ref{teff} is that without either lepton
or charged kaon identification for comparison with case (a)-(c).
The outputs of the neural-network studies are shown in Figures \ref{feff}
(a), (b), and (c) for the three cases described so far, respectively.
In case (c), a hadron collider B-factory without a RICH was assumed.
In this case, if
the $B^0$ vertices are well identified, only
charged tracks from the $B^0$ can be selected.
The exact four-vector of the $B^0$ was unknown, however for simplicity,
the four-vector is assumed to be
known exactly.

In case (a), the improvement was only 10\%  as seen in Table \ref{teff}.
However, in cases (b) and (c),
significant improvements were observed. In case (b), the RICH performance
was almost recovered. In case (c), the performance of the weighted charge
method was close to half that of the case of perfect kaon identification.
In a practical case, the four-vector of $B^0$ is unknown.
One must therefore
use the transverse momentum with respect to the $B^0$-jet axis
instead of three-vectors at the CMS frame. This will decrease the performance
of this tagging.

As can be
seen from Figures 3 (a), (b), and (c), the peaks at around 0.1 and 0.9
are considered to be those for prompt particles.
The broad structures at around 0.35 and 0.65 are considered to be
other contributions
such as high-momentum kaons.

%statistics of gold plated events
In a real experiment, there are two ways to tune the cut parameters.
One is to tune them by a Monte-Carlo simulation, and
the other to obtain
$\epsilon_{eff}$
using more than 1000 fully reconstructed $B^0$ events.
Here, a fully-reconstructed $B^0$ event means an event where one of the
$B^0$'s is reconstructed exclusively. There, however, is an ambiguity
in the $B^0-\overline{B^0}$ oscillation in the determination of
the B flavor of the other $B^0$.
The other method is to obtain more than 10000
fully reconstructed
events for tuning in order to avoid fragmentation model ambiguities.
Also, at a B-factory, one expects $>1\times 10^7$ $B^0 \overline{B^0}$
events per year.
Therefore, the $B^0$ reconstruction efficiency of 0.1\% will
give 20000 reconstructed events. Both methods are considered to be valid.

%importance in B decay-mode study including theory
In order to establish this method, the $B^0$ decay-mode studies at
CLEO are important. This method strongly relies on the
fragmentation model of $(W)\rightarrow q\bar{q}\rightarrow $hadrons.
Also, a theoretical consideration
concerning this nature would
be of great benefit.

%discussion at total tagging efficiency (improvement)
In the neural-network studies, the simplest parametrization was used, i.e.,
three-vectors and the charges of the four highest-momentum particles.
It may be more effective if a new parametrization is found.
For example in B-factory cases, neutral particles can be used for jet
identifications.

%conclusion
To conclude, a new method to identify the B flavor of $B^0$ mesons
in a CP violation measurement at the asymmetric B-factory was
developed. This method does not require either lepton
or charged-kaon identification. The tagging efficiency, probability
for incorrect tagging, and
effective tagging efficiency were obtained to be 43.1, 18.3, and
17.3\%, respectively.
This effective tagging efficiency is significantly large
compared with that of lepton tagging. In the case with
charged-kaon tagging, however, the gain is small.
This method will be helpful especially at an initial stage of
the B-factory where such particle identification devices as RICH cannot be
fully operated.

%ack(none)
I appreciate discussions with members of the BELLE collaboration.

\end{document}